\documentclass{ifacconf}
\usepackage{xcolor}
\usepackage{bm}
\usepackage{enumitem}

\usepackage{amsmath}
\usepackage{amssymb}
\usepackage{graphicx}      
\usepackage{natbib}        
\begin{document}
\begin{frontmatter}

\title{A new safety-guided design methodology to complement model-based safety analysis for safety assurance \thanksref{footnoteinfo}} 

\thanks[footnoteinfo]{This work is sponsored by NASA Grant 80NSSC19K1702, with Natasha Neogi and Jon Holbrook acting as technical monitors.}

\author[First]{Minghui Sun} 
\author[Second]{Cody H. Fleming} 

\address[First]{University of Virginia, 
   Charlottesville, USA (ms3yq@virginia.edu).}
\address[Second]{Iowa State University, 
   Ames, USA (flemingc@iastate.edu)}

\begin{abstract}  
With the rapid advancement of Formal Methods, Model-based Safety Analysis (MBSA) has been gaining tremendous attention for its ability to rigorously verify whether the safety-critical scenarios are adequately addressed by the design solution of a cyber-physical human system. However, there is a gap. If specific safety-critical scenarios are not included in the given design solution (i.e., the model) in the first place, the results of MBSA cannot be trusted for safety assurance. To tackle this problem, we propose a new safety-guided design methodology (called STPA+) to complement MBSA. Inspired by STPA, STPA+ treats a system as a control structure, which is particularly fit for systems with complex interactions between human, machine, and automation. Three methods are developed in STPA+ to tackle the possible omissions of safety-critical scenarios caused by incorrectly defined safety constraints, improperly constrained process model, and inadequately designed controller. In this way, STPA+ directly derives an adequately defined design solution as the input to an MBSA verification program and bridges the gap between current MBSA approaches and safety assurance. 
\end{abstract}


\end{frontmatter}

\section{Introduction}
Cyber-physical human systems are pervasive in today's world. For example, the modern commercial airplane heavily depends on the collaboration of human pilot and cockpit automation to safely and efficiently transport passengers or cargo from Point A to Point B. Certifying a commercial airplane requires by law evidence that all the possible safety-critical scenarios that the airplane will encounter in the actual operation are adequately addressed, which is especially challenging for new aircraft entrants that have novel features (e.g., 737MAX) and future advanced air mobility concept. Similar challenges abound in many other cyber-physical human system domains, such as automated driving \cite{kramer2020identification} and nuclear power plant \cite{ahn2015development}.

\begin{figure}[h!]
\begin{center}
\includegraphics[width=0.31\textwidth]{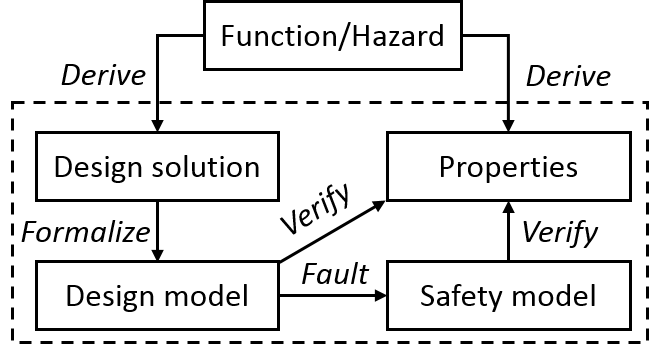}
\vspace{-6pt}
\caption{Given the function and the hazard, the mainstream MBSA approaches (in the dotted box) contribute to safety assurance by verifying the given design solution against the given properties.} 
\label{fig:mbsa}
\end{center}
\end{figure}

With the rapid advancement of Formal Methods, Model-based Safety Analysis (MBSA) \cite{sun2021defining} has been gaining tremendous attention for its ability to rigorously verify whether the safety-critical scenarios are properly addressed by a design solution. Many MBSA approaches have been developed over the past two decades. The prominent ones include AADL \cite{feiler2006architecture}, AltaRica \cite{prosvirnova2014altarica}, and HipHops \cite{kabir2018dynamic}, which according to \cite{bozzano2015safety}, are the only three languages that ``have matured beyond the level of research prototypes''. In general (Fig.\ref{fig:mbsa}), these mainstream MBSA approaches start with a given design solution by formalizing it into a design model in the target languages; then faults and failures are modeled to achieve a safety model; finally, verification is conducted to see whether the given properties (functional or safety, deterministic or probabilistic) are satisfied.

However, there is a gap. Even assuming MBSA (or formal verification in a more general sense) can exhaustively check all the possible scenarios captured by the design solution, there are still open challenges. First, verification is already challenging for complex hybrid systems. More importantly, if certain safety-critical scenarios are not included or considered in the given design solution (the left shoulder of Fig.\ref{fig:mbsa}), or the properties do not correctly reflect the hazard (the right shoulder of Fig.\ref{fig:mbsa}), the results of MBSA cannot be fully trusted for safety assurance.

In fact, such a gap has already been identified by one of FAA's reports \cite{butka2015advanced} that systems that pass formal verification can still be incorrect. \cite{leveson2020you} further argues ``formal verification can only show consistency between two formal models'' and ``cannot be used as a validation of acceptable
safety''. Instead, Leveson proposes Systems Theoretic Process Analysis (STPA). STPA is known for its ability to identify hazardous scenarios associated with control \cite{steck2018methodological} and define safety properties from the hazard \cite{hobbs2021risk}. Furthermore, STPA is also designed for analyzing systems with complex human-automation interactions \cite{fleming2016early}, while the more traditional Fault Tree Analysis, as prescribed by ARP 4761 \cite{4761} for aviation certification, does not even consider humans or software in the fault tree, let alone analyzing the interaction between them. All these traits make STPA an ideal alternative to bridge this gap for cyber-physical human systems.

However, STPA has two weaknesses. First, STPA has to model an existing design solution into a control structure, which ultimately relies on the ``art of modeling'' and opens the window for inconsistency between the design model and the STPA model. Second, STPA provides less support in identifying the causes of the hazardous scenarios associated with control \cite{leveson2013stpa}, which is why many approaches \cite{kramer2020identification, steck2018methodological} only use STPA to identify the safety-critical scenarios for the system as a whole, similar to, e.g., HAZOP. However, it is equally important to decompose these scenarios into more refined scenarios that the design solution can directly act upon. According to a recent review \cite{zhang2021finding} on identifying the safety-critical scenarios in the Automated Driving community, no work is found that identifies the safety-critical scenarios to support defining the design solution.

Therefore, this paper proposes a new \textbf{safety-guided design methodology}, called STPA+, based on STPA to (1) derive the design solution directly and (2) provide better methodological support to identify and then address the safety-critical scenarios that the design solution should directly act upon. Because the safety-critical scenarios caused by failure have been addressed by MBSA (the bottom arrow in Fig.\ref{fig:mbsa}), STPA+ only focuses on the safety-critical scenarios that happen without the presence of component-level failures. In fact, there is an ISO standard under development (ISO/PAS 21448, SOTIF) that specifically focuses on this type of scenarios. As a result, STPA+ strengthens the two shoulders of Fig.\ref{fig:mbsa}, and ``STPA+MBSA'' provides more comprehensive safety assurance for cyber-physical human systems. 

Finally, because STPA+ does not need an existing design solution to identify the safety-critical scenarios that a design solution should directly act upon, STPA+ can also be potentially used to design a run-time safety monitor for systems of which one has limited control over the operation, such as a legacy system or an AI-based system. The run-time monitor may function as a modular safety feature to capture hazardous scenarios and make decisions when the original system fails to act. 

\section{Overview of STPA+}
\begin{figure}[h!]
\begin{center}
\includegraphics[width=0.48\textwidth]{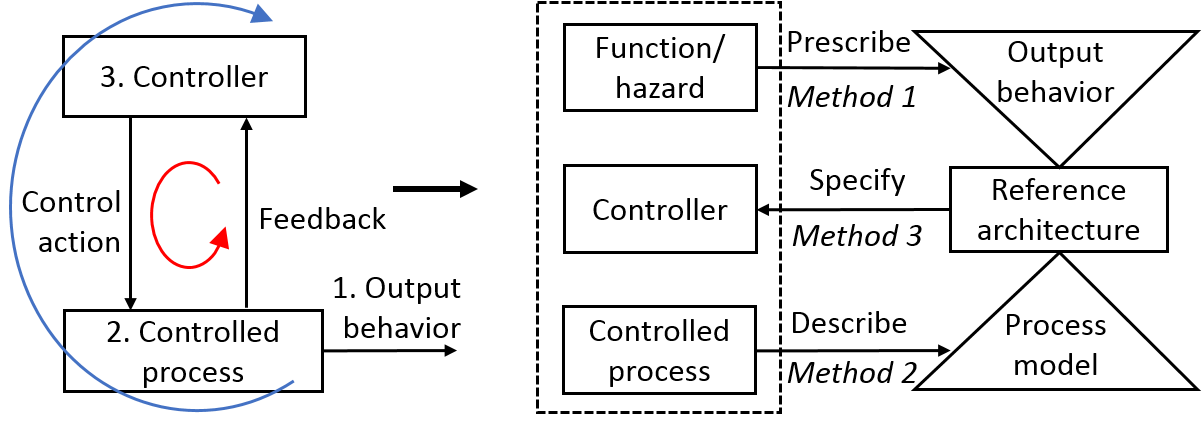}
\vspace{-6pt}
\caption{An overview of STPA+. Left is the control structure that operates counter-clockwise and is designed clockwise; right is the proposed STPA+ comprised of three methods. } 
\label{fig:stpa+}
\end{center}
\end{figure}

STPA+ shares the same system-theoretic view as STPA: a system can be abstracted as a (hierarchical) control structure (the left of Fig.\ref{fig:stpa+}) regardless of a human controller, automation, or a team of both, which makes it an ideal candidate to design cyber-physical human systems. 

A control structure operates in the following order (the red circle of Fig.\ref{fig:stpa+}): (1) the controller observes the controlled process, (2) the controller issues control actions to manipulate the evolution of the controlled process, (3) the output behavior of the controlled process (called \textbf{output behavior} hereafter) achieves the functional goal and avoids the hazard.

Designing a control structure follows the opposite direction (the blue circle of Fig.\ref{fig:stpa+}), which translates into the three methods of STPA+ below (the right of Fig.\ref{fig:stpa+}).
 \begin{itemize}[leftmargin=*]
     \item Method 1: Translating the function/hazard into constraints on the output behavior. Because these constraints are the ``desired end results'' of the control structure as a whole, they are \emph{prescriptive constraints}. 
     \item Method 2: Properly constraining a model of the controlled process. Because these constraints represent what the controlled process can possibly do, they are \emph{descriptive constraints}.
     \item Method 3: A reference architecture for the controller to issue the right control action at the right time so that the constraints from both directions will be respected. 
 \end{itemize}

\section{Deriving the prescriptive constraints (Method 1)}
The prescriptive constraints are defined on the output behavior. The output behavior comprises three elements: a start time $st$, a stop time $sp$, and the trajectory evolution in $[st, sp]$. Accordingly, constraints must be derived from the hazard for all three elements. Safety constraints in the current practice are mostly the restrictions on the trajectory evolution. For example, a minimal distance from the terrain is defined to constrain the descent of an airplane when it is descending. Such constraint on the trajectory evolution while the output behavior is ongoing is called \textbf{performance constraint} in this paper. However, the performance constraint does not prescribe when the output behavior should start or stop. The lack of constraints on the start/stop time may lead to the omission of safety-critical scenarios, especially for time-sensitive output behaviors. Therefore, the question is, \emph{how to derive the safety constraints on the start/stop time from the hazard}. 
\begin{figure}[h!]
\begin{center}
\includegraphics[width=0.45\textwidth]{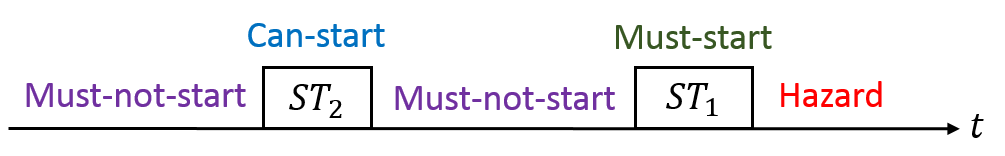}
\vspace{-6pt}
\caption{An example of the constraints on the start time.} 
\label{fig:time}
\end{center}
\end{figure}

First, consider the start time $st$. For example (Fig.\ref{fig:time}), a time-sensitive output behavior is safe to start only within $ST_1$ and $ST_2$, which divides the timeline into different sections with different meanings. The output behavior ``must start'' when the timeline comes to $ST_1$, otherwise the hazard will surely happen after $ST_1$; the output behavior ``can start'' when the timeline comes within $ST_2$, because if not, it can still start in $ST_1$; the output behavior ``must not start'' in the rest of the sections. Clearly, there are three types of constraints for the start time:
\begin{itemize}[leftmargin=*]
    \item Must-start time window $mst\_T$: the output behavior must start within this time window, otherwise the hazard will happen after the time window expires. 
    \item Must-not-start time window $nst\_T$: the output behavior must not start within this time window, otherwise the hazard will happen immediately. 
    \item Can-start time window $cst\_T$: It is safe whether the intended output behavior starts or not when the time comes to $cst\_T$. Mathematically, the can-start time window is complementary to the $mst\_T$ and $nst\_T$, and can be calculated in (\ref{eq:cst}), where $\mathcal{T}$ is the time before $mst\_T$ expires. 
    \begin{equation}\label{eq:cst}
        cst\_T=\mathcal{T}\cap\neg(mst\_T\cup nst\_T)
    \end{equation}
\end{itemize}

Second, we reason about the time windows. For the must-start condition, how can the hazard happen if the intended output behavior has not even started yet? That must be caused by the output behavior temporally immediately before the intended output behavior (called \textbf{in-behavior} hereafter) violating its own performance constraints (denoted as $pc^i$). For the must-not-start condition, the hazard is caused by the started intended output behavior violating its own performance constraints (denoted as $pc$). Therefore, the must-start time window is defined to avoid the in-behavior from stopping at the wrong time to violate $pc^i$; the must-not-start time window is to avoid the intended output behavior from starting at the wrong time to violate $pc$. 
 \begin{figure}[h!]
\begin{center}
\includegraphics[width=0.47\textwidth]{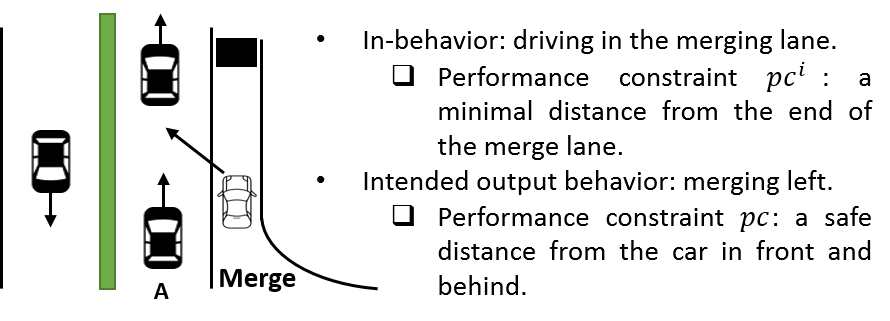}
\vspace{-6pt}
\caption{The merging example. } 
\label{fig:merge}
\end{center}
\end{figure}

Taking merging into the traffic for example, with the in-behavior and the intended output behavior and their respective performance constraints defined in Fig.\ref{fig:merge}. The ``must-start'' time window is determined based on the time left before the car breaches the minimal distance from the end of the merge lane (i.e., the in-behavior violating $pc^i$). The ``must-not-start'' time window is determined based on the time when there is no safe opening to merge into the traffic (i.e., the intended output behavior violating $pc$). 
\begin{figure}[h!]
\begin{center}
\includegraphics[width=0.45\textwidth]{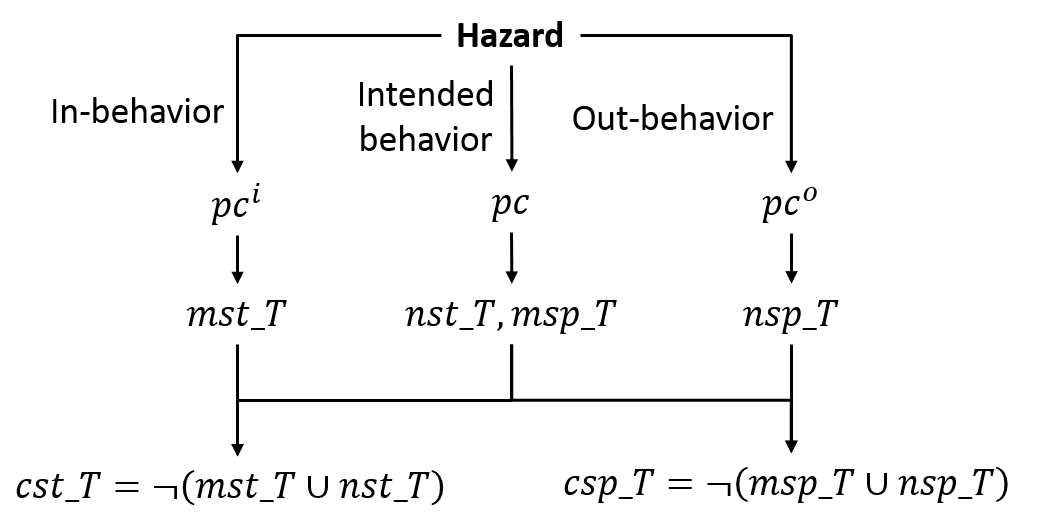}
\vspace{-8pt}
\caption{Deriving the constraints on the start/stop time. The \textbf{out-behavior} is the output behavior temporally immediately after the intended output behavior. } 
\label{fig:m1}
\end{center}
\end{figure}

 As a result, the constraints on the start/stop time can be summarized in Fig.\ref{fig:m1}. The must-start, must-not-start and can-start time window ($mst\_T,nst\_T$ and $cst\_T$) are derived based on the performance constraints of the in-behavior $pc^i$ and the performance constraints of the intended output behavior $pc$. Similarly, the must-stop, must-not-stop and can-stop time window ($msp\_T,nsp\_T$ and $csp\_T$) are derived based on $pc$ and the performance constraints of the out-behavior $pc^o$.  Eventually, the prescriptive constraints of the output behavior $y(t)$ where $t\in[st,sp]$ can be defined in (\ref{eq:m1}), and $Y(t),ST$ and $SP$ are the final prescriptive constraints. 
 \begin{equation}\label{eq:m1}
     \begin{cases}
      y(t)\in Y(t)\subseteq pc, \text{ where }  t\in[st,sp]\\
      st\in ST\subseteq mst\_T\cup cst\_T\cap\neg nst\_T\\
      sp\in SP\subseteq msp\_T\cup csp\_T\cap\neg nsp\_T\\
     \end{cases}
 \end{equation}

\section{Deriving the descriptive constraints (Method 2)}
Method 2 is to constrain a model of the controlled process properly. An incorrect model can lead to hazardous scenarios that will escape the scrutiny of any model-based analysis. General System Theory \cite{gst} provides a generic construct $(u(t),x(t),p)\xrightarrow{f}(\Dot{x},y(t))^T$ to model the controlled process, where $u$ is the control, $p$ is the set of parameters, $x$ is the system state, and $y$ is the output. Since the basic construct can only be determined case-by-case,the question is, \emph{how to properly constrain a given construct $(f,u, x, p,\Dot{x},y)$ for a controlled process?}  

\begin{figure}[h!]
\begin{center}
\includegraphics[width=0.25\textwidth]{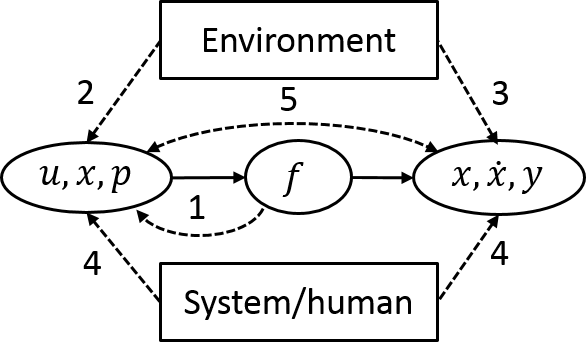}
\vspace{-6pt}
\caption{A map to identify the constraints of the controlled process. Each dotted arrow is a type of constraints from the source to the destination. } 
\label{fig:m2}
\end{center}
\end{figure}

Fig.\ref{fig:m2} provides a map to identify the constraints. Each dotted arrow may yield both the \textbf{constraints} on $(u, x, p,\Dot{x},y)$ and the associated \textbf{assumptions} on the environment and the system/human.

First, $f$ represents concrete mechanisms (engineered or natural) of the actual process. It may only be able to process a finite set of the inputs and hence have constraints on the inputs (Arrow 1).

Second, the controlled process operates on actual components. For those that are out of the design scope (i.e., the designer has no control), we call them ``environment''. Components within the design scope are the ``system'' to be built or further refined, or ``human'' to be trained. For the environment, it may have a set that includes all the possible inputs (Arrow 2) and another set in which all the possible outputs must be included (Arrow 3). For the system/human, it is always subject to finite design/manufacture/natural capacities that may yield constraints on both inputs and outputs (Arrow 4).

Third, $(u, x, p,\Dot{x},y)$ is internally constrained by $f$. Therefore, the constraints on $(u, x, p,\Dot{x},y)$ also need to satisfy $f$ mathematically (Arrow 5). 

Finally, each constraint may have assumptions on the environment and the system/human for the constraint to be valid in the first place. Therefore, constraints and assumptions should always appear in pairs, and any isolated constraint or assumption must be justified.    

\begin{figure}[h!]
\begin{center}
\includegraphics[width=0.3\textwidth]{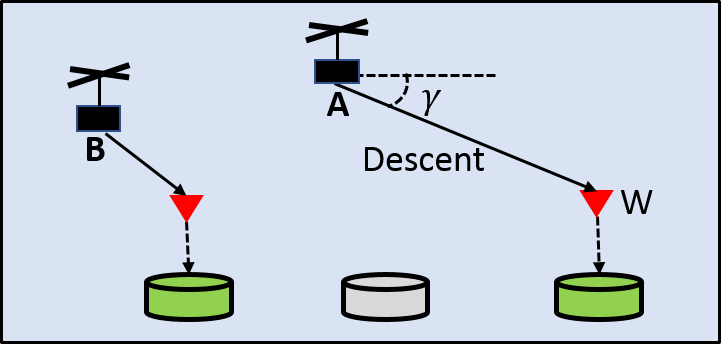}
\vspace{-6pt}
\caption{An example of eVTOL descent. The grey landing pad is backup for emergency use only.} 
\label{fig:descent}
\end{center}
\end{figure}

Therefore, all the \textbf{constraint-assumption pairs} (denoted as \{constraint$\vert$ assumption on system/human$\vert$ assumption on environment\}) need to be identified based on Fig.\ref{fig:m2} to properly constrain the model of the controlled process. For example (Fig.\ref{fig:descent}), Vehicle A (an eVTOL) is instructed to descend to Point W with a constant descent angle $\gamma$. We treat $\gamma$ as a system state and derive its constraint-assumptions pair based on Fig.\ref{fig:m2}. Note that Arrow 2 and Arrow 5 do not apply to this example.   

\noindent \textbf{Arrow 1}. $f$ here is to maintain a constant descent angle, which can only be achieved when $\gamma$ is within $\Gamma_1$ if the battery level is more than $BL$ percent and the crosswind is less than $CW$ knots. The constraint-assumption pair hence is $\{\Gamma_1|BL|CW\}$. \newline
\textbf{Arrow 3}. To avoid interference with the adjacent landing pad, the vertiport (i.e., the environment) requires $\gamma$ to be greater than $\Gamma_3$. However, when the backup landing pad is activated in an emergency, the vehicle will need to descend at a greater angle. Therefore, the nominal vertiport operation is an assumption on the environment. The constraint-assumption hence is $\{\Gamma_3|NA|Nom\}$.\newline
\textbf{Arrow 4}. The vehicle can afford a descent angle within $\Gamma_4$ if the local elevation is lower than $EL$ and the vehicle payload weighs less than $PD$. The constraint-assumption hence is $\{\Gamma_4|PD|EL\}$.

As a result, the constraints of $\gamma$ are $\Gamma_1\cap\Gamma_3\cap\Gamma_4$, and the associated assumptions are $BL\wedge CW\wedge Nom\wedge PD\wedge EL$.

Finally, the same procedure is applied to each element of $(u, x, p,\Dot{x},y)$ for the constraint-assumption pair. Eventually, the descriptive constraints can be represented in (\ref{eq:gst}), where $AS$ is the set of the assumptions.
\begin{equation}\label{eq:gst}
    \begin{cases}
     (u(t),x(t),p)\xrightarrow{f}(\Dot{x},y(t))^T\\
     (u, x, p,\Dot{x},y)\in (U, X, P,\Dot{X},Y)\\
     AS= AS_x\wedge AS_u\wedge AS_p\wedge AS_{\Dot{x}}\wedge AS_y
    \end{cases}
\end{equation}

\section{Reference architecture for a safe controller (Method 3)}
Mathematically, the next task is to design a controller to find the control
actions to satisfy the constraints from (\ref{eq:m1}) and (\ref{eq:gst}). However, a safe controller in an actual system is more than a mathematical problem solver. It is a decision-maker that constantly monitors
the environment and the controlled process, watches out for hazards, reacts to
changes, and adjusts the decisions in real-time. To design such a safe controller
is to identify all the safety-critical scenarios and design mechanisms into the controller to address them accordingly.

First, a controller in the most general sense, must make the following three decisions. 
\begin{enumerate}[leftmargin=*]
    \item Decide prescriptive constraints: In a dynamic environment, what is safe/unsafe usually changes in real-time. The controller must adjust the prescriptive constraints accordingly. 
    \item Decide control reference: Because the controller needs control references to generate the control action, this decision is to decide the control references from the prescriptive constraints. 
    \item Decide control action: This decision is akin to the control algorithm in traditional feedback control: generating control actions from given control references.
\end{enumerate}
Taking ``lane merging'' in Fig.\ref{fig:merge} for example, Decision (1) is to find the road section and time window to merge safely by monitoring the traffic; Decision (2) is to decide the specific position and time to merge; Decision (3) is to operate the steering and gas to execute the merge.

\begin{figure}[h!]
\begin{center}
\includegraphics[width=0.49\textwidth]{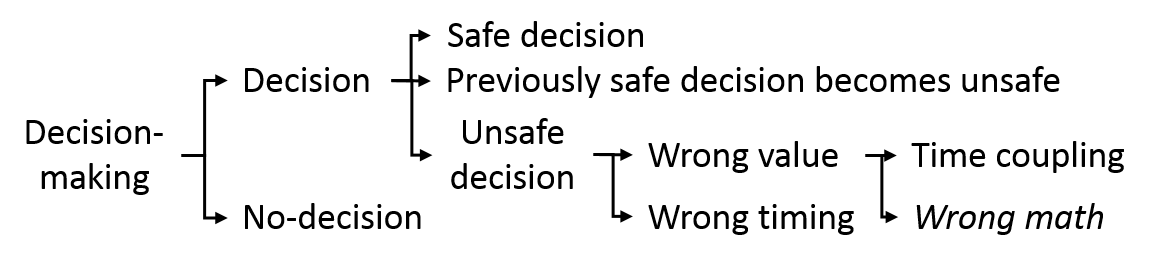} 
\vspace{-22pt}
\caption{The five types of unsafe scenarios (not involving failure) when making a decision. The ``Previously safe decision becomes unsafe'' scenario is addressed as ``\emph{previously safe}'' for short. } 
\label{fig:scenario}
\end{center}
\end{figure}
Second, regardless of the decision being made, there are five types of unsafe scenarios (without failure/fault) when making a decision (Fig.\ref{fig:scenario}). (i) \emph{No-decision} refers to the scenarios where no decision can be found. (ii)\emph{Previously safe} refers to the scenarios where an already generated decision becomes unsafe due to change. (iii)\emph{Unsafe timing} refers to the scenarios where the decision is made too late, so much so the timing constraints in (\ref{eq:m1}) cannot be met as the operation of the rest of the system also takes time. (iv)\emph{Time coupling} refers to the scenarios where the result of a decision is affected by the time delay of the system, i.e., time and value are coupled, such as the phenomenon of Pilot-induced Oscillations. (v)\emph{Wrong math} refers to the scenarios where the algorithm to calculate the decision is wrong or inadequate. Most works in theoretical control or formal methods are to get the math right. Therefore, we \emph{do not} consider this scenario in the reference architecture. All the unsafe scenarios above are scenarios, if not appropriately addressed, can cause a hazard.

Third, we consider how the safety-critical scenarios explained above (except the``wrong math'') can happen to each of the three decisions that a controller makes. We will continue using the ``lane merging'' as a running example for the first two decisions. 

\noindent \textbf{(1) Decide prescriptive constraints.}
This decision observes information from outside the controller and generates the prescriptive constraints in (\ref{eq:m1}) based on the procedure of Method 1. \emph{Time coupling} is not applicable.

\noindent\emph{No-decision:} There are internal conflicts between the constraints, e.g. the entire must-start time window is included by the must-not-start time window (i.e. $mst\_T\subseteq nst\_T$). For example, the traffic is dense, and no opening can be found to merge safely till the end of the merging lane.   

\noindent\emph{Previously safe:} The prescriptive constraints become stricter due to the change in the environment, or the prescriptive constraints are violated by the output behavior due to uncertainty. For the former scenario, the traffic becomes tighter after the safe merging section and time window are found, which requires the merging section and time window to be adjusted accordingly. For the latter, the car merges at a point out of the selected merging section because a hole in the road disrupts the heading, making the car dangerously close to the traffic. 

\noindent\emph{Unsafe timing:} The prescriptive constraints are decided too late. For example, the driver finds the safe merging section just before the car hits the end of the merge lane, which is too late because the car needs more time to initiate and accomplish the merge. 

\noindent \textbf{(2) Decide control reference.} This decision generates the control reference so that the output behavior will satisfy the prescriptive constraints in (\ref{eq:m1}), and the trajectory evolution will respect the descriptive constraints in (\ref{eq:gst}). Note that the trajectory evolution of the controlled process can be determined by the control reference, given the dynamics, the control algorithm, and the initial condition.  

\noindent\emph{No-decision:} The controller cannot find a control reference to satisfy the prescriptive constraints and the descriptive constraints at the same time. For example, the only available merging opening requires an acceleration rate that exceeds the car's designed performance. Hence, no target merging point is found. 

\noindent\emph{Previously safe:} The prescriptive or descriptive constraints can be violated due to the change of the constraints or the deviation of the actual controlled process from the planned trajectory evolution.
\begin{itemize}[leftmargin=*]
\item The prescriptive constraints \emph{are going to} be violated by the output behavior. For example, the traffic suddenly becomes tighter (i.e., the prescriptive constraints change) when the car is actively merging. The car is now predicted to be unsafely close to the traffic after the merge, and therefore should stop merging. 
\item The descriptive constraints \emph{are going to} be violated by the trajectory evolution. For example, the car enters a road section with a slower speed limit (i.e., the descriptive constraints change) when it is speeding up to merge. The eventual merging speed that is acceptable for the previous speed limit is predicted to violate the new speed limit. 
\item The descriptive constraints \emph{being} violated by the current states of the controlled process. For example, the car enters a downhill while it is speeding up to merge, making the car extra faster than planned (i.e., deviation from the planned trajectory evolution). As a result, the speed limit is violated before the merge. The car must slow down immediately instead of speeding up to merge. 
\end{itemize}

\noindent\emph{Unsafe timing:} The control reference may be issued too late without considering the fact that the execution of the rest of the system takes time, which eventually leads to a violation of the timing constraint in (\ref{eq:m1}). 

\begin{figure}[h!]
\begin{center}
\includegraphics[width=0.37\textwidth]{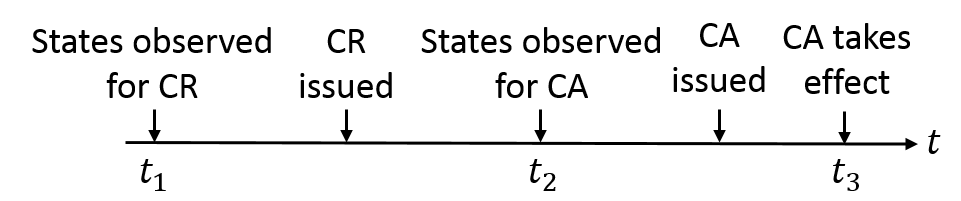}
\vspace{-9pt}
\caption{The event sequence from generating the control reference at $t_1$ to generating the control action at $t_2$, to the controlled process actively seeking the control reference at at $t_3$. CR and CA stand for control reference and control action respectively. } 
\label{fig:delay}
\end{center}
\end{figure}

\noindent\emph{Time coupling:} In the most general sense (Fig.\ref{fig:delay}), the control reference starts to be generated at $t_1$ and the controlled process starts to seek the control reference at $t_3$. When such an event sequence operates fast enough, the states observed at $t_1$ can be considered the same as the actual states at $t_3$. However, when the time delay between $t_1$ and $t_3$ is unneglectable, the initial states used to generate the control reference should not be the states observed at $t_1$, but rather the predicted states at $t_3$ based on the observation at $t_1$. The coupling between the time delay and the control reference must be addressed properly to avoid the ``wrong value'' scenario in Fig.\ref{fig:scenario}.

\noindent \textbf{(3) Decide control action.} This decision is akin to the control algorithm in the theoretical control, to generate control action based on given control reference. 

\noindent\emph{No-decision:} As long as the control reference is found, the control action is also available because Decision 2 makes sure the descriptive constraints can always be satisfied. However, if the generated control reference needs to be updated, only the control reference after $t_c+t_3-t_1$ can be updated ($t_c$ is the current time) because of the time delay $t_3-t_1$ in Fig.\ref{fig:delay}. As a result, the control action within $[t_c,t_c+t_3-t_1]$ must be updated in a way that the constraints (prescriptive and descriptive) can be satisfied at the same time. No-decision scenario is hence when such control action cannot be found. 

\noindent\emph{Previously safe:} For systems that operate at a slower pace, the control actions may be issued a certain time before being executed. This scenario means the issued control actions may lead to a violation of the constraints (prescriptive and descriptive) due to either the change of the constraints or the deviation of the controlled process from the predicted trajectory evolution. For example, an air traffic controller recalls the previous instruction when the spacing constraints change (i.e., constraints change) for the airspace ahead; or the airplane deviates from its planned course due to a wind gust (i.e., deviation), causing the air traffic controller to change its previous instruction. 

\noindent\emph{Unsafe timing:} The control action may be issued too late without considering the fact that the execution of the controlled process takes time.

\noindent\emph{Time coupling:} Similar to the previous decision. When the time delay between $t_2$ and $t_3$ (Fig.\ref{fig:delay}) is unneglectable, the time delay between $t_2$ and $t_3$ must be accounted for when generating the control action. 

Finally, the scenarios identified above can already be used to examine whether an existing controller design is safe. However, STPA+ takes one step further by defining sub-decisions within each decision to address all the safety-critical scenarios. Although the details are out of the scope of this paper, we define 33 main sub-decisions, 12 enabling sub-decisions, and their interactions as a reference architecture. The reference architecture works like a \emph{class} (as in Object-Oriented Programming). In the specific design applications, engineers only need to instantiate the reference architecture and tailor it for their own problems, yielding a controller that has all the safety-critical scenarios (without component-level failure) automatically addressed. We will explain the details of the reference architecture in an extension of this work. 

\section{Conclusion}
MBSA has gained tremendous traction over the past two decades due to its ability to rigorously prove whether all the safety-critical scenarios captured in the design solution are addressed correctly. However, to apply MBSA for certification of safety-critical systems, one still needs to demonstrate that the model used in MBSA includes all the possible safety-critical scenarios that the actual system will encounter in actual operation. Current MBSA approaches are not equipped for this task.

To tackle this problem, we propose a new safety-guided design methodology, called STPA+, to complement MBSA for safety assurance. STPA+ treats a system as a control structure, which is particularly fit for systems with complex interactions between human, machine, and automation. Furthermore, STPA+ comprises three methods. Method 1 is to derive safety constraints from the hazard to make sure the safety constraints adequately reflect the hazard under study. Method 2 is to define the model of the controlled process to make sure the model is properly constrained both explicitly and implicitly. Method 3 is to define a safe controller (human, automation, or a team of both) by providing a reference architecture to make sure the controller defined based on the reference architecture has all the safety-critical scenarios (without component-level failure) addressed. In summary, Method 1 strengthens the right shoulder of Fig.\ref{fig:mbsa}; Method 2\&3 strengthens the left shoulder of Fig.\ref{fig:mbsa}. Together, ``STPA+MBSA'' provides a strong and comprehensive argument for the safety assurance of a cyber-physical human system.  

In the future, we will demonstrate specifically how STPA+ can address the concerns in ISO/PAS 21448 (Safety of the intended functionality). Moreover, because STPA+ does not require knowledge about the design solution to identify the safety-critical scenarios that a controller must directly act on, we will apply STPA+ to design a run-time monitor as a modular safety feature to assist the human/AI-based controller or the automation controller of a legacy system to capture and act on hazardous scenarios when necessary.

\bibliography{ifacconf}             
                                                   







\end{document}